\documentclass[prl,nofootinbib,superscriptaddress,twocolumn]{revtex4}
\usepackage[a4paper,left=1.5cm,right=1.5cm,top=3cm,bottom=3cm]{geometry}

\usepackage{amssymb,amsmath,amsfonts}
\usepackage[dvipsnames]{xcolor}
\usepackage{graphicx}
\usepackage{longtable}
\usepackage{verbatim}
\usepackage{color}
\usepackage{mdframed}
\usepackage{soul}
\usepackage{bm}
\usepackage{amsfonts,amssymb,mathrsfs,amsmath,esint}
\usepackage{slashed, cancel}
\usepackage{framed}
\usepackage{mdframed}
\usepackage{simplewick} 
\allowdisplaybreaks

\usepackage{latexsym}
\usepackage{graphicx}
\graphicspath{{./Figures/}}
\usepackage[dvipsnames]{xcolor}
\usepackage{booktabs}
\usepackage{datetime}
\newdateformat{mydate}{\THEDAY{ }\monthname[\THEMONTH]{ }\THEYEAR}

\usepackage{tikz}
\usepackage{color}
\usepackage{framed}
\usepackage{hyperref}
\hypersetup{colorlinks, citecolor=bluscuro, linkcolor=black, urlcolor=bluscuro}
\definecolor{rossos}{cmyk}{0,1,1,0.55}
\definecolor{bluscuro}{rgb}{0.15, 0.2, .85}
\definecolor{bluchiaro}{cmyk}{1,.3,0.,0.1}

\graphicspath{{./images/}}

\newcommand{\be}{\begin{equation}}
\newcommand{\ee}{\end{equation}}
\newcommand{\bea}{\begin{eqnarray}}
\newcommand{\eea}{\end{eqnarray}}
\newcommand{\beq}{\begin{equation}}
\newcommand{\eeq}{\end{equation}}

\def\beqa{\begin{eqnarray}}

\def\eeqa{\end{eqnarray}}

\def\lsim{\mathrel{\rlap{\lower4pt\hbox{\hskip0.5pt$\sim$}}
    \raise1pt\hbox{$<$}}}         
\def\gsim{\mathrel{\rlap{\lower4pt\hbox{\hskip0.5pt$\sim$}}
    \raise1pt\hbox{$>$}}}         

\usepackage[normalem]{ulem}
\usepackage{soul}

\begin{document}

\title{Signatures of Primordial Gravitational Waves \\on the Large-Scale  Structure of the Universe}

\author{Pritha Bari}
\address{Dipartimento di Fisica e Astronomia ``G. Galilei",
Universit\`a degli Studi di Padova, via Marzolo 8, I-35131 Padova, Italy}
\address{INFN, Sezione di Padova,
via Marzolo 8, I-35131 Padova, Italy}
\email{pbari@pd.infn.it,\\ angelo.ricciardone@pd.infn.it,\\ nicola.bartolo@pd.infn.it,\\daniele.bertacca@pd.infn.it,\\sabino.matarrese@pd.infn.it}

\author{Angelo Ricciardone}
\address{Dipartimento di Fisica e Astronomia ``G. Galilei",
Universit\`a degli Studi di Padova, via Marzolo 8, I-35131 Padova, Italy}
\address{INFN, Sezione di Padova,
via Marzolo 8, I-35131 Padova, Italy}

\author{Nicola Bartolo}
\address{Dipartimento di Fisica e Astronomia ``G. Galilei",
Universit\`a degli Studi di Padova, via Marzolo 8, I-35131 Padova, Italy}
\address{INFN, Sezione di Padova,
via Marzolo 8, I-35131 Padova, Italy}
\address{INAF - Osservatorio Astronomico di Padova, I-35122 Padova, Italy}
\author{Daniele Bertacca}
\address{Dipartimento di Fisica e Astronomia ``G. Galilei",
Universit\`a degli Studi di Padova, via Marzolo 8, I-35131 Padova, Italy}
\address{INFN, Sezione di Padova,
via Marzolo 8, I-35131 Padova, Italy}
\address{INAF - Osservatorio Astronomico di Padova, I-35122 Padova, Italy}

\author{Sabino Matarrese}
\address{Dipartimento di Fisica e Astronomia ``G. Galilei",
Universit\`a degli Studi di Padova, via Marzolo 8, I-35131 Padova, Italy}
\address{INFN, Sezione di Padova,
via Marzolo 8, I-35131 Padova, Italy}
\address{INAF - Osservatorio Astronomico di Padova, I-35122 Padova, Italy}
\address{Gran Sasso Science Institute, I-67100 L’Aquila, Italy}

\date{\today}

\begin{abstract}

We study the generation and evolution of second-order energy-density perturbations arising from primordial gravitational waves. Such ``tensor-induced scalar modes'' approximately evolve as standard linear matter perturbations and may leave observable signatures in the Large-Scale Structure of the Universe. 
We study the imprint on the matter power-spectrum of some primordial models which predict a large gravitational-wave signal at high frequencies. 
This novel mechanism in principle allows us to constrain or detect primordial gravitational waves by looking at specific features in the matter or galaxy power-spectrum, thereby allowing to probe them on a range of scales unexplored so far. 

\end{abstract}

\maketitle
\paragraph{\label{sec:level1}Introduction} 
Inflation in the Early Universe \cite{Guth:1980zm} plays a crucial role in the standard model of Cosmology, as it provides the seeds for structure formation through 
scalar (energy density) perturbations originating from quantum vacuum oscillations of the scalar field driving the accelerated universe expansion. At the same time, inflation produces tensor (gravitational-wave) perturbations by quantum fluctuations of the metric tensor. It is usually believed that only scalar modes feed structure formation, while primordial Gravitational Waves (GW) contribute to Cosmic Microwave Background (CMB) temperature anisotropies and polarization, while not affecting the Large-Scale Structure (LSS) of the Universe. The only considered exceptions to this rule comes from the so-called ``tensor fossils'', where large amplitude long-wavelength GW couple to scalar modes giving rise to specific anisotropic signatures in the LSS \cite{Masui:2010cz, Jeong:2012df, Dimastrogiovanni:2014ina}, and from indirect effects of GW in the LSS clustering and shear~\cite{Kaiser:1996wk,Jeong:2012nu,Schmidt:2012nw} or galaxy shapes~\cite{Biagetti:2020lpx}.

In this paper we explore a novel mechanism for generating matter-density perturbations based upon the non-linear evolution of primordial tensor modes. This mechanism was first proposed and analyzed in \cite{Tomita1971,Tomita1972, Matarrese:1997ay}. These tensor-induced scalar modes are statistically independent of standard adiabatic density perturbations, so that the overall matter power-spectrum is merely the sum of the ones from the two components. Due to the fact that gravitational waves are frozen on super-horizon scales, they can source our second order perturbations only on sub-horizon scales, so our effect does not produce CMB anisotropies on large scales, unlike the linear matter perturbation. Here we consider scales which entered the Hubble radius after matter-radiation equality. A more detailed analysis where smaller scales are included will be presented in a separate paper~\cite{Barietal2}.

Our main result is that primordial GW can lead to observable effects in the matter power-spectrum. In particular, we focus here on models of inflation where the linear tensor power-spectrum is either blue-tilted or endowed with a Gaussian bump, as it happens, e.g., in Axion Inflation models~\cite{Namba:2015gja,Thorne:2017jft,Dimastrogiovanni:2016fuu}. Such models, which are also good candidates to be probed by next generation GW interferometers, such as LISA or ET~\cite{Bartolo:2016ami,Maggiore:2019uih}, leave an observable imprint on LSS. This opens the possibility to constrain or to detect primordial GWs through future LSS surveys, such as Euclid \cite{Euclid:2019clj}, DESI \cite{DESI:2018ymu}, SPHEREx \cite{2016arXiv160607039D}, SKA \cite{SKA:2018ckk}, Roman Space Telescope \cite{2021arXiv211103081R} and Vera Rubin Observatory (LSST)\cite{2020arXiv200907653V}.

\paragraph{\label{sec:new}Tensor-induced scalar modes}
Nowadays, the use of linear perturbation theory (e.g. \cite{Mukhanov:1990me, mukhanov2005physical}) is well justified for very large scales and as long as mainly the (matter) power-spectrum is considered. On the other hand, higher-order perturbation theory \cite{Matarrese:1992rp, Matarrese:1993zf,2021arXiv210710283D} (or more sophisticated resummation techniques \cite{Matarrese:2007wc, Matsubara:2007wj,Senatore:2014via,Vlah:2015zda}) is needed, as soon as one extends the analysis to higher-order correlators (such as the bispectrum or trispectrum) or aims at describing LSS formation on mildly non-linear scales and/or in connection with galaxy bias schemes.
In this context, one of the main focuses, recently revived, has been the scalar perturbations as seeds of second-order tensor ones, for the obvious reason that they are the dominant ones at linear order. Discussion of scalar-induced gravitational waves can be found in various works, such as \cite{Matarrese:1997ay,Matarrese:1996pp, Mollerach:2003nq,Baumann:2007zm,Inomata:2019yww,Yuan:2019fwv, Ananda:2006af,Kohri:2018awv,Domenech:2020kqm,Espinosa:2018eve,Saito:2008jc,2019arXiv190912708G} (see also~\cite{2021Univ....7..398D} for a review).

In this paper, we present the opposite case, i.e. we look for the signature of GW on cosmic structures. 
Detecting the primordial GW background is one of the major goals of cosmology, pursued both through CMB polarization data \cite{BICEP2:2018kqh, LiteBIRD:2020khw,2020arXiv200812619T} and, after the recent groundbreaking detection of astrophysical GW \cite{LIGOScientific:2016aoc}, also at future interferometers \cite{Caprini:2019pxz, Campeti:2020xwn,Flauger:2020qyi}. According to the mechanism studied in what follows, GW produced in the early Universe (see, e.g. \cite{Maggiore:1999vm,Guzzetti:2016mkm,Watanabe:2006qe,BICEP2:2018kqh}) can source scalar perturbations upon re-entering the horizon \cite{NoteX}. By studying their effect on the matter perturbations, we also aim at providing a new way to constrain the tensor-to-scalar perturbation ratio on scales where we have poor constraints. 

In our analysis, we consider a spatially flat Friedmann-Lema\^itre-Robertson-Walker (FLRW) background metric perturbed up to second order, $d s^2= a^2(\eta)\left[-d\eta^2+\gamma_{ij}({\bf x},\eta) d x^i d x^j\right]$; here $\eta$ is the conformal time, and $a(\eta)$ the scale-factor. In our analysis we restrict ourselves to include collision-less cold dark matter plus a cosmological constant; these simplifications allow us to perform our calculations in the 
synchronous and comoving gauge, which, because of the absence of pressure gradients, can also be made time-orthogonal \cite{Kodama:1985bj}. The conformal spatial metric $\gamma_{ij}$ contains linear scalar and tensor modes (linear vector modes are not considered here, as they would decay in an expanding Universe). At second order one has scalar-driven scalar, vector and tensor perturbations, second-order terms mixing linear scalar and tensors (i.e., ``tensor fossils'', as mentioned above),  tensor-induced
vector and tensor modes \cite{Matarrese:1997ay}, and finally the tensor-induced scalar modes, which we are most interested in here (see also \cite{Zhang:2017hiu,Carrilho:2015cma}). Since the latter are statistically independent of linear scalar modes, we are allowed to deal separately with them, recovering the effect of standard density perturbations at the end. Hence, we take only tensor modes at first order and only scalar ones at second order: 
$\gamma_{ij} = (1 - \phi^{(2)})\delta_{ij} + (1/2) D_{ij}\chi^{||(2)} + \chi_{ij}^{(1)T}$ 
where $\chi_{ij}^{(1)T}$ are the linear tensor modes (GW), 
$D_{ij} = \partial_i \partial_j - (1/3) \nabla^2 \delta_{ij}$, 
$\phi^{(2)}$ and $\chi^{||(2)}$ are tensor induced scalar metric perturbations. 

We get the fluid deformation tensor by subtracting the isotropic background expansion from the covariant derivative of the four-velocity $\theta^\mu_\nu = a u^\mu_{;\nu}-\mathcal{H}\delta^\mu_\nu$.
Choosing comoving observers yields a huge advantage, as it keeps the fluid four-velocity orthogonal to the constant-time spatial hypersurface (described by $\gamma_{ij}$), so that our $\theta^\mu_\nu$ is purely spatial, coinciding with the extrinsic curvature of
constant-time spatial hypersurfaces, $\theta^i_j= -K^i_j=\gamma^{ik}\gamma'_{kj}/2$, with a prime denoting differentiation w.r.t. conformal time.
Our main equations are the Raychaudhuri and continuity equations \cite{Matarrese:1997ay, Bruni:2013qta}
\begin{align}\label{eq:ray}
    \theta'+\mathcal{H}\theta +\theta^i_j \theta^j_i+4\pi G a^2 \overline{\rho}_m \delta&=0\,,\\\label{eq:del}
     \delta'+\big(1+\delta\big)\theta&=0\,,
\end{align}
where $\theta$ is the {\it peculiar} volume expansion scalar, $\mathcal{H} \equiv a'/a$, $\overline{\rho}_m$ the mean energy density of the matter component and $\delta$ its density contrast, $\delta=(\rho_m-\overline{\rho}_m)/\overline{\rho}_m$.

We write the density perturbation as $\delta = \delta^{(1)}+ \delta^{(2)}/2$, and similarly for $\theta$. Now, from \eqref{eq:ray} and \eqref{eq:del} at second order we get
\begin{align}
  {\theta'}^{(2)}+\mathcal{H}\theta^{(2)}+2\theta^{(1)i}_j \theta^{(1)j}_i+4\pi G a^2 \overline{\rho}_m \delta^{(2)}&=0\\
   \delta^{'(2)}+2  \delta^{(1)} \theta^{(1)}+ \theta^{(2)}&=0\, .
\end{align}
We combine these to get ( $\chi_{ij}^{(1)T}=\chi_{ij}$ from here on):
\begin{equation}\label{evolgen}
    {\delta}^{(2)''}+\mathcal{H}{\delta}^{(2)'}-4\pi G a^2 \overline{\rho}_m \delta^{(2)}=\frac{1}{2} {\chi'}^{ij}\chi'_{ij}\,.
\end{equation}
As expected, the LHS of this equation coincides with the evolution equation for the linear density contrast, but a source term appears, which is quadratic in the tensor perturbation modes. Remembering that the GW energy density is given by $\rho_{GW}= (1/32\pi Ga^2) \langle {\chi'}^{ij}\chi'_{ij}\rangle$, it is clear that this is the quantity sourcing $\delta^{(2)}$ in \eqref{evolgen}.
\paragraph{\label{sec:level3}Density contrast}
The homogeneous and sourced solutions of \eqref{evolgen} are, respectively,
\begin{align}\label{homogen}
     \delta^{(2)}_h= {}& c_1\big(\bm{x}\big)D_+(\eta)+c_2\big(\bm{x}\big)D_-(\eta)\,,\\
\begin{split}\label{inhomogen}
    \delta^{(2)}_s={}& D_+(\eta)\int_0^\eta d\tilde{\eta} \frac{ D_-(\tilde{\eta})}{W(\tilde{\eta})}\frac{1}{2}{\chi'}^{ij}\chi'_{ij}\\-& D_-(\eta)\int_0^\eta d\tilde{\eta}\frac{ D_+(\tilde{\eta})}{W(\tilde{\eta})}\frac{1}{2}{\chi'}^{ij}\chi'_{ij}\,,
\end{split}
\end{align}
where $D_+$ and $D_-$ are the linear growing and decaying homogeneous solutions, and $W(\eta)\equiv D_-(\eta)D_+'(\eta)-D_+(\eta)D_-'(\eta)$
is the Wronskian.
%
From \eqref{inhomogen} we can see that our density contrast, though derived at second order, evolves in time just like the linear one. 

Here we focus on scalar modes entering the horizon during matter domination (we will include the effects of dark energy later on).
In order to compute the power-spectrum we move to Fourier space and write 
\begin{equation}
    \chi_{ij}\big(\bm{x},\eta\big)= \sum_{\sigma} \int \frac{d^3\bm{k}}{\big(2\pi\big)^3} e^{i\bm{k}\cdot\bm{x}} \chi_\sigma \big(\bm{k}, \eta\big)\epsilon^\sigma_{ij}\big(\bm{\hat{k}}\big) \,,
\end{equation}
where $\epsilon^\sigma_{ij}(\bm{\hat{k}})$ are the polarisation tensors [i.e. $\epsilon^\sigma_{ij}(\bm{\hat{k}}){\epsilon^{\sigma' ij}}(\bm{\hat{k}})= 2\delta_{\sigma \sigma'}$] for the two GW polarizations $\sigma=+, \times$, and $\chi_\sigma (\bm{k}, \eta)$ is the GW mode function which sources the scalar perturbations. Having in mind scales which entered the Hubble radius in matter domination, we can make use of the following tensor transfer function~\cite{Watanabe:2006qe}
$\chi_\sigma (\bm{k}, \eta)= A_\sigma (\bm{k})[{3j_1(k\eta)}/{k\eta}]$,
where $j_1$ is the spherical Bessel function of order one and $A_\sigma \big(\bm{k}\big)$ is a stochastic zero-mean random field characterized by the following auto-correlation function,  
\begin{equation}\label{aa}
     \langle A_{\sigma_1}\big(\bm{k}_1\big) A_{\sigma_2}\big(\bm{k}_2\big)\rangle= \big(2\pi\big)^3 \delta^3\big(\bm{k}_1+\bm{k}_2\big) \delta_{\sigma_1 \sigma_2} \frac{2\pi^2}{k_1^3}\Delta^2_{\mathcal{\sigma}}\big(k_1\big)\,,
\end{equation}
$\Delta^2_{\mathcal{\sigma}}(k)$ being the power-spectrum for each GW polarisation. 

The time integral in \eqref{inhomogen} can be split into two parts, from the end of inflation to matter-radiation equality, and from equality to the observation time.
The growing and decaying solutions for density perturbations, as well as the transfer function for the source GWs, should be appropriately chosen for the respective integrals. 
In the first, radiation-dominated part, linear density perturbations 
involve two modes, $D_+=\ln{\eta}$ and $D_-={\rm const.}$, whereas the GW transfer function behaves as $j_0(k\eta)$. While, in the matter era $D_+=\eta^2$ and $D_-=\eta^{-3}$ and the GW transfer function is proportional to $j_1(k\eta)$. However, as we will explain later, the contribution from the first part is negligible compared to the second one.

The power-spectrum of the stochastic GW background depends on the specific mechanism by which it was generated. In this paper we mainly focus on inflationary models which are characterized either by a monochromatic spectrum, or by a blue tensor spectrum or by a Gaussian bump. 
Moreover, we consider parity-invariant
mechanisms of production of GWs, so we expect that both polarisations give the same effect, leaving for a future analysis the possibility to consider parity breaking early Universe models.
\paragraph{\label{sec:level4}Power-spectrum}  
Our final goal is to study the present-day matter power-spectrum, taking into account the effect from this tensor-induced perturbation, and explore how to infer such an effect from its imprint on the LSS. 
Following the discussion before, we can evaluate the time integral purely in the matter era, extending the lower limit of time integration down to 0.
According to the definition of matter power-spectrum $\Delta^2_{\mathcal{\delta}}(k)$, \begin{equation}\label{dd}
    \langle \delta^{(2)}\left(\bm{k},\eta\right)\delta^{(2)}\big(\bm{k}',\eta\big)\rangle= \big(2\pi\big)^3 \delta^3\big(\bm{k}+\bm{k}'\big) \frac{2\pi^2}{k^3}\Delta^2_{\mathcal{\delta}^{(2)}}\big(k\big)\,,
\end{equation}
and using Eq. \eqref{aa}, we get the following expression
 \begin{multline}\label{MDp}
    \Delta^2_{\mathcal{\delta}^{(2)}}\big(k\big)=\frac{k^3}{2\pi }\sum\limits_{\sigma,\sigma'}\int d^3\bm{k}_2 \frac{\Delta^2_{\sigma'}\big(k_2\big)\Delta^2_{\sigma}\big(|\bm{k}-\bm{k}_2|\big)}{k_2^3 |\bm{k}-\bm{k}_2|^3}\\
   \times f\big(k,k_2,\theta\big)\\
   \times \left[\frac{\eta^2}{10}\int_0^\eta \frac{d\tilde{\eta}}{\tilde{\eta}}\Big(\frac{3j_1\big(k_2 \tilde{\eta}\big)}{k_2\tilde{\eta}}\Big)'\Big(\frac{3j_1\big(|\bm{k}-\bm{k}_2| \tilde{\eta}\big)}{|\bm{k}-\bm{k}_2|\tilde{\eta}}\Big)'\right.\\-\left. \frac{1}{10\eta^3}\int_0^\eta d\tilde{\eta}\, \tilde{\eta}^4\Big(\frac{3j_1\big(k_2 \tilde{\eta}\big)}{k_2\tilde{\eta}}\Big)'\Big(\frac{3j_1\big(|\bm{k}-\bm{k}_2| \tilde{\eta}\big)}{|\bm{k}-\bm{k}_2|\tilde{\eta}}\Big)'\right]^2\,,
\end{multline}
where we have defined $f(k,k_2,\theta)$ to be the following contraction of the polarisation tensors
\begin{multline}
 f\big(k,k_2,\theta\big)\equiv  \\ \sum\limits_{\sigma,\sigma'}\epsilon^{\sigma'}_{ij}\big(\bm{\hat{k}}_2\big)\epsilon^{\sigma ij}\big(\bm{\widehat{k-k_2}}\big)\epsilon^{\sigma'}_{kl}\big(\bm{-\hat{k}_2}\big)\epsilon^{\sigma kl}\big(\bm{\widehat{-k+k_2}}\big)\,.\nonumber
\end{multline}
Here $\theta$ is the angle between $\bm{\hat{k}}$ and $\bm{\hat{k}}_2$. In the convolution, and all the expressions from hereon, $\bm{k}$ always corresponds to the induced scalar modes, and $\bm{k}_2$, $\left(\bm{k}-\bm{k}_2\right)$ correspond to the source GW modes.
Here, we confine ourselves to the standard practice of dealing with the growing mode only, which we believe to be sufficient. GW modes are frozen outside the horizon, so from \eqref{inhomogen}, we can see that in \eqref{MDp} the contribution will come after horizon entry, when GW start oscillating with an amplitude damped by a factor $\sim 1/a$. As we are considering scalar modes crossing the horizon during matter domination, we can safely switch on our sourcing at $\eta=0$, since at that early time our modes are super-horizon, and hence are not triggered. Moreover, since after entering the horizon, GWs decay away, we consider $\eta \rightarrow \infty$ as an upper bound of the time integral instead of putting the exact age of the Universe, as most of the contribution will come from around the time of horizon-entry anyway.

To solve such integrals, it is useful to work with the variables $x=k_2/k$, $y=|\bm{k}-\bm{k}_2|/k$, and use the dimensionless time variable $\tau=k\tilde{\eta}$ \cite{Bartolo:2007vp}. Considering only the growing-mode term in the square brackets of \eqref{MDp} we get
\begin{multline}
\label{asol}
    \Delta^2_{\mathcal{\delta}^{(2)}}\big(k\big)= \\\frac{81k^4\eta^4}{100} \int_0^\infty dx \int_{|x-1|}^{x+1} dy \big(xy\big)^{-2}  f\big(x,y\big) \Delta_{\sigma}^2\big(x k\big)\Delta_{\sigma}^2\big(yk\big)\\
    \times\left[\int_0^\infty \frac{d \tau}{\tau^3}j_2\big(x \tau\big)j_2\big(y \tau\big)\right]^2\,, 
    \end{multline}
where the function $f$ in terms of $x$ and $y$ now reads
\begin{multline}
    f\big(x,y\big)= \frac{1}{16x^4y^4}\Big[x^8+\big(y^2-1\big)^4+4x^6\big(7y^2-1\big)\\+4x^2\big(y^2-1\big)^2\big(7y^2-1\big)+x^4\big(6-60y^2+70y^4\big)\Big]\,.
\end{multline}
There are two important points to stress here: one is the wave-vector integration domain that now, in terms of $x$ and $y$, is given by \cite{Espinosa:2018eve} $(x+y)\ge 1\, \land $  $(x+1)\ge y \,\land$   $(y+1)\ge x$; the second is the time integration domain, that has been extended to $\eta\rightarrow\infty$, for the reasons explained above. This allows to solve the integral analytically in terms of hyper-geometric functions. We have indeed checked that the contribution from $\eta$ to infinity is negligible compared to the contribution from $0$ to $\eta$.

Let us stress here that, although our density contrast modes strictly lie in the matter domination regime, in calculating \eqref{asol}, we are integrating over the whole frequency range of GW modes, using the appropriate transfer functions shown in \eqref{MDp}. This is despite the fact that the transfer functions in \eqref{MDp} are valid
only for tensor modes entering the horizon during matter domination.
However, we have checked that tensor modes entering during radiation
domination constitute a negligible contribution of the resulting
signal. 
\paragraph{Results for different GW sources}
\label{sec:level5}
The spectrum of the GW background depends on the details of the generation mechanism. Besides the standard vacuum oscillations
of the gravitational field during inflation, there are many other well-motivated early Universe scenarios which can produce a large GW background also at small scales. Within inflationary mechanisms this is the case, e.g., of models where the inflaton is coupled to gauge (i.e., $U(1)$ or $SU(2)$) fields~\cite{Barnaby:2010vf,Sorbo:2011rz}, or models where space-diffeomorphisms are broken during inflation \cite{Endlich:2012pz, Bartolo:2015qvr,Ricciardone:2016lym}. Also, primordial black holes (PBHs), formed via the gravitational collapse of small-scale curvature perturbations, are a powerful source of second-order GWs \cite{Bartolo:2018evs,Saito:2008jc}. Regarding post-inflationary mechanisms, a strong GW signal can be produced by topological defects \cite{Vilenkin:1981bx,Figueroa:2012kw}, or phase-transitions \cite{Kosowsky:1991ua,Kamionkowski:1993fg}. As we will show in the coming section, all the models that have a large monopole GW signal, can source the density perturbation affecting in this way the matter power-spectrum.

\begin{figure}
    \centering
    \includegraphics[width=8.6cm]{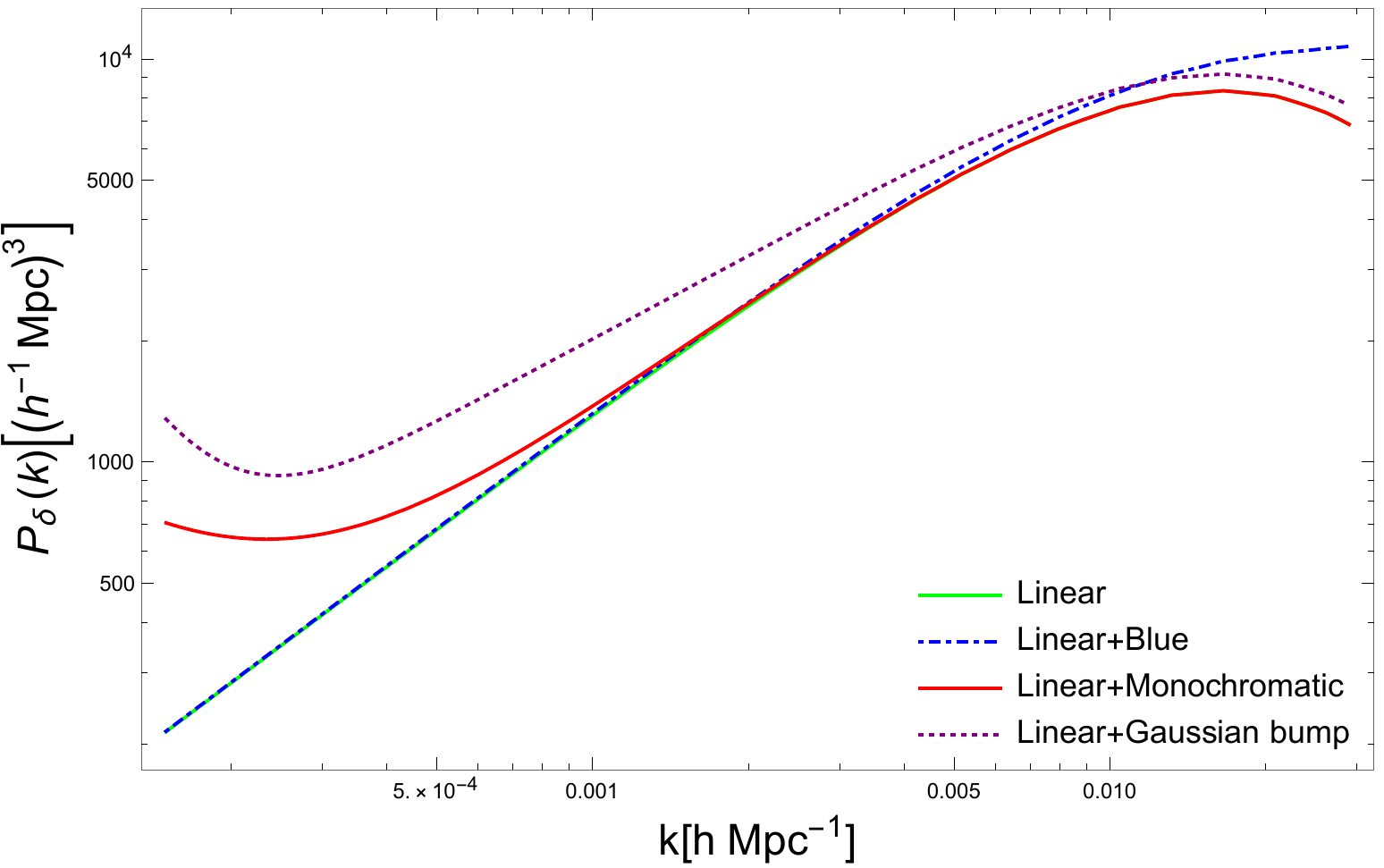}
    \caption{\it Impact of different GW power spectra on the  matter power-spectrum: i) blue-tilted~$\left(A_T=1.26\times 10^{-10}, n_T = 0.32, k_*=k_{\rm{CMB}}=0.01 {\rm{Mpc}^{-1}}\right)$, ii) monochromatic~$\left(A_T=10^{-5}, k_*=0.008 {\rm{Mpc}^{-1}}\right)$, iii) Gaussian bump~$\left(A_T=10^{-5}, \sigma = 2, k_p = 0.04 \rm{Mpc}^{-1}\right)$. The value of $h$ is fixed at $0.68$\cite{Planck:2018vyg}.}
    \label{Allinone}
\end{figure}

So, now we are going to quantify the impact on the matter power-spectrum \eqref{asol} for different input GW signals. The power-spectrum for individual polarisation modes is related to the GW power-spectrum by the relation,  $\Delta^2_{\sigma}(k)= (1/2)\Delta^2_T(k)$. It is important to stress that, moving away from CMB scales, we have less stringent bounds on the amplitude of the GW spectrum. This allows us to choose larger values for their amplitude. 
%
\subparagraph{Power-law spectrum} 
As a first benchmark signal, we consider a power-law spectrum, which is typical of many single-field inflationary models~\cite{Guzzetti:2016mkm}
\begin{align}\label{flat}
     \Delta^2_T\big(k_2\big)&=A_T \left(\frac{k_2}{k_*}\right) ^{n_T}\,.
\end{align}
Here $A_T$ is the amplitude at a given reference or pivot scale $k_*$, and $n_T$ is the tensor spectral index. Usually $A_T$ is translated in terms of the tensor-to-scalar ratio $r$, defined as the ratio between the tensor and scalar
power-spectrum amplitudes at $k_*$, $r_{k_*}=A_T/A_S$. 
Standard single-field,
slow-roll inflationary models predict a nearly scale-invariant  power-spectrum on super-horizon scales (with $n_T =-2 \epsilon$, with $\epsilon$ being the usual slow-roll parameter). On CMB scales the tensor-to-scalar ratio is strongly constrained by the latest Planck data,
$r_{0.01} < 0.066$ (at 95\% Confidence Level (C.L.), PLANCK TT,TE,EE+lowE+lensing+BK15+LIGO$\&$Virgo2016), constraining $-0.76<n_T<0.52$ \cite{Planck:2018jri}. 
We consider the case of a blue-tilted tensor power-spectrum, choosing $n_T = 0.32$, which is still within the range of values allowed by present and future GW interferometers~\cite{Bartolo:2016ami,Planck:2018jri}, and fix the GW power-spectrum amplitude at $A_T=r A_s=0.06 \times 2.1 \times 10^{-9}=1.26\times 10^{=10}$. We can observe in Fig. \ref{Allinone}, that the effect starts to be relevant, especially on relatively large $k$.
\subparagraph{Monochromatic Spectrum} A useful case-study is a monochromatic tensor spectrum, which can be regarded as an approximation of a spectrum with a sharp peak 
\begin{equation}\label{mon}
    \Delta^2_T\left(k_2\right)=A_T \delta\left(\ln{\frac{k_2}{k_*}}\right)\,.
\end{equation}
Typical models that predict such a spectrum are   \cite{Cai:2020ovp}.
In this case the form of the power-spectrum can be found analytically and it reads \newpage
\begin{eqnarray}\label{mono}
   \Delta^2_{\mathcal{\delta}^{(2)}}(k)& =& 4\times 10^{-5}\big(k\,\eta_0\big)^4 A_T^2 \nonumber\\
   &\times&  \left(\frac{8k_*^2}{k^2}+\frac{k^6}{16k_*^6}-\frac{k^4}{2k_*^4}+3\frac{k^2}{k_*^2}-8\right)\Theta(2 k_{*}-k)\,,\nonumber\\
\end{eqnarray}
where the condition $k<2k_*$ comes from momentum conservation.
\subparagraph{Gaussian-bump spectrum} A well-motivated early Universe scenario which predicts a large and characteristic amplitude for the GW spectrum consists in a GW signal endowed with a large Gaussian bump.
An example where we see this kind of bump is an axion field coupled to SU(2) gauge fields as spectator fields besides the inflaton \cite{Namba:2015gja,Thorne:2017jft}.
Such models 
result in GWs which are amplified at the same level as the scalar perturbation and so they are possible targets both for CMB B-mode observations~\cite{Thorne:2017jft} and interferometers~\cite{Bartolo:2016ami, Garcia-Bellido:2016dkw}.
Therefore, for our purposes, one can study their signatures on the matter power-spectrum as another way to test or constrain such early Universe scenarios. 
The tensor power-spectrum is characterized by the following Gaussian bump
\begin{equation}
 \Delta^2_T \big(k_2\big)=A_T 
 e^{-\frac{1}{2\sigma^2}\ln^2\big(\frac{k_2}{k_p}\big)}\, .
\end{equation}

The impact on the matter power-spectrum is visible in Fig. \ref{Allinone}, where the effect of the Gaussian bump reflects also on the shape of the  matter power-spectrum.
In Fig.~\ref{Allinone} we report the standard matter power-spectrum and the one including the effects computed in this paper. The dimensional power-spectrum of matter we show in the plot is related to the dimensionless $\Delta^2_{\mathcal{\delta}^{(2)}}(k)$, derived from \eqref{asol}, through the relation  $P_{\mathcal{\delta}^{(2)}}(k)=(2\pi^2/k^3)\Delta^2_{\mathcal{\delta}^{(2)}}(k)$. Some of these effects are similar/degenerate with other cosmological observable like non-Gaussianity, projection effects, kinetic dipole, finger of the observer, and wide-angle effects (e.g., see \cite{Jeong:2011as, Raccanelli:2016avd, Abramo:2017xnp, 2021arXiv210700351B, 2021arXiv210813424Y}), so it is important to find also other peculiar signatures.

\begin{figure}
       \includegraphics[width=8.6cm,height=5.5cm]{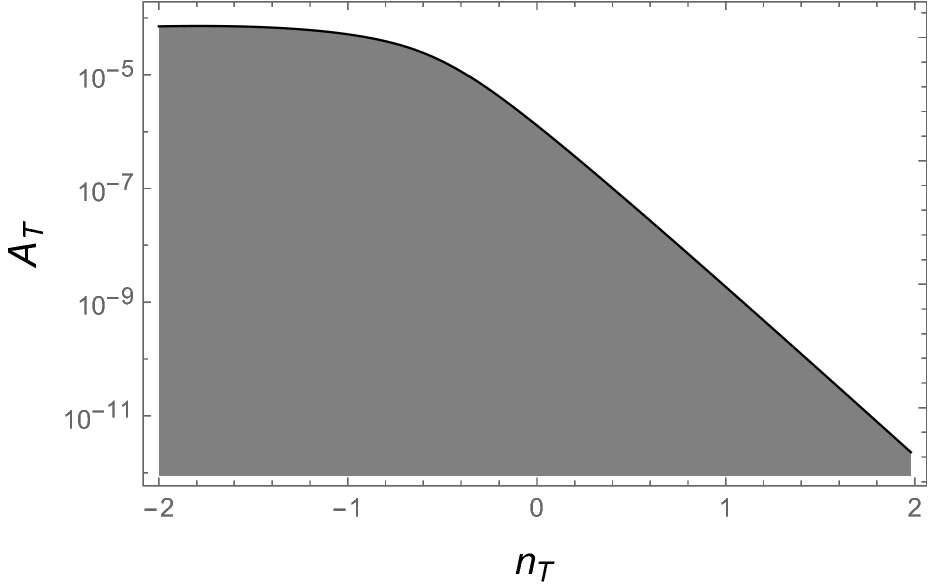}
    \caption{\it The region of parameter space  for $n_T$ and $A_T$ where the power-spectrum of the GW-sourced density perturbation mode with the wavenumber $k=0.006 \rm{Mpc}^{-1}$ obeys a $4\%$ error bound on the linear matter power-spectrum for a power-law GW source. The grey region shows the allowed range of the parameters  assuming the mentioned  error uncertainty.}
       \label{Parameter}
\end{figure}

 We can see that different kinds of GW sources result in extra contributions to the matter power-spectrum which are comparable to, and exceeding the linear power-spectrum, in different ranges of wave-numbers. 
 We calculate \eqref{asol} as a function of $A_T$ and $n_T$ for a power-law GW spectrum \eqref{flat}, and considering a 4$\%$ error bound w.r.t. the linear matter power-spectrum at  $k=0.006 \rm{Mpc}^{-1}$, we obtain the parameter space shown in Fig. \ref{Parameter}. The grey region shows the allowed range for $n_T-A_T$ accounting for the mentioned error. 
 
 \paragraph{Including the effect of the dark energy}\label{sec:level6} All of our analyses so far have been done under the assumption that after matter-radiation equality, the Universe is dominated by cold dark matter only. 
 Choosing an $\Omega_m$ different from 1 results in different linear growth factor $D_+$ (in \eqref{homogen} or \eqref{inhomogen}), which suppresses the matter growth.
 The fitting formula for the growth suppression factor for linear density perturbation is given in \cite{peacock_2010, 1992ARA&A..30..499C}.
 Using $\Omega_m=0.32$ \cite{Planck:2018vyg}, we find our suppressed 
 power-spectrum to be $P_{\mathcal{\delta}^{(2)}}\left( z=0, \Omega_m =0.32\right) \simeq 0.59 P_{\mathcal{\delta}^{(2)}}\left( z=0, \Omega_m=1 \right)$.
Fig. \ref{Allinone} already includes the dark energy component. 
\paragraph{Early time evolution}\label{sec:level7} In the previous sections, we have discussed the induced matter perturbation modes entering the horizon during matter domination. As anticipated, the contribution from the modes which enter before matter-radiation equality is negligible compared to the former one. Here, we briefly discuss what we may have to face in the era when radiation was the dominant component. Second-order perturbations in synchronous gauge for the scalar-tensor and tensor-tensor couplings are discussed in \cite{Zhang:2017hiu} and \cite{Wang:2019zhj} respectively for the matter and radiation dominated stages, although their full matter 
power-spectra were not studied. 

At early times, the Universe consists of a mixture of radiation and pressure-less matter. Since there are two matter components now, we would not have the advantage of having the fluid four-velocity tensor orthogonal to the spatial hypersurface anymore. Instead, we have to use the conservation equations for both components, and the Einstein equations. Since the radiation and cold dark matter components interact only gravitationally, their energy–momentum tensors satisfy their conservation laws separately. The standard way of dealing with this scenario is to consider two phases: first, very early times when the gravitational potential is solely determined by radiation 
and determines matter perturbations, and second, when the matter perturbation grows significantly towards the equality time, and dominantly contributes to the potential.  In this case
we would get the Meszaros' equation for the sub-horizon evolution of matter perturbation in a Universe filled with radiation and matter \cite{dodelson2021modern,Hu_1996,Meszaros:1974tb,1975A&A....41..143G}, but now with a source term. In order to know the full non-linear evolution at second order, we then need to solve the inhomogeneous equation. This is a treatment we leave for a future work.  
\paragraph{\label{sec:level8}Discussion and conclusion} In this paper we have analyzed the new effect of ``tensor-induced scalar modes'' on the present day matter power-spectrum.  We have found that a large GW 
power-spectrum can leave an imprint on the matter power-spectrum. 
There are two important features of our second-order matter-perturbation: first, we do not have a contribution on super-horizon scales, unlike the linear matter perturbation, and as a result, our effect does not produce CMB temperature anisotropy on large scales. Second, our matter density contrast completely mimics the linear one on the sub-horizon scales. In that sense, it can be considered as a linear perturbation, sourced by gravitational radiation, vanishing outside the horizon. A distinguishing feature would be its high intrinsic non-Gaussianity, which we intend to explore in the future. 
We also showed the parameter space for the tensor spectral index versus the GW amplitude, accounting for the uncertainties on the measurements of the matter power-spectrum.
This signature can be useful both for detecting and constraining GWs in a novel way 
on a range of scales on which we currently have very poor information, and to increase the accuracy in the estimation of matter 
power-spectrum.
\paragraph{Acknowledgments.}
\noindent
P.B. and S.M. sincerely thanks Serena Giardiello for her help in clarifying some concepts. A.R. thanks Stefano Anselmi, Mauro Pieroni and Lorenzo Valbusa Dall'Armi for useful comments and discussions. N.B., D.B and S.M. acknowledge support from the COSMOS network (www.cosmosnet.it) through the ASI (Italian Space Agency) Grants 2016-24-H.0 and 2016-24-H.1-2018. A.R. and P.B.~acknowledges funding from Italian Ministry of Education, University and Research (MIUR) through the ``Dipartimenti di
eccellenza'' project Science of the Universe.
\vskip 0.1cm
\bibliographystyle{ieeetr}
\bibliography{name.bib}


\end{document}